\documentclass[letterpaper,journal,10pt]{IEEEtran}

\usepackage{cite}

\usepackage{graphicx}
\usepackage{multirow}
\usepackage{flushend}

\usepackage[cmex10]{amsmath}
\usepackage{amssymb}
\usepackage{nicefrac}

\usepackage{fixltx2e}

\usepackage{booktabs}

\usepackage{ifpdf}
\ifpdf
\usepackage[draft,pdfborder={0 0 0}]{hyperref}
\fi

\usepackage{bm}
\usepackage[varg]{txfonts}
\let\mathbb=\varmathbb
\DeclareSymbolFont{letters}{OML}{ztmcm}{m}{it}

\usepackage{tikz,pgf}
\usepackage{pgfplots}
\usetikzlibrary{shapes,positioning,arrows,decorations.markings,fit,calc,patterns}

\usepackage{algorithm}
\usepackage{algorithmicx}
\usepackage{algpseudocode}
\algrenewcommand{\algorithmiccomment}[1]{\textbf{//}#1}

\usepackage{subfig}

\usepackage{color}

\DeclareMathOperator*{\argmax}{arg\,max}

\newcommand{\mvec}[1]{\bm{#1}}
\newcommand{\est}[1]{\hat{u}_{#1}}
\newcommand{\estvec}[2]{\hat{\bm{u}}_{#1}^{#2}}
\newcommand{\prob}[1]{\text{Pr}[#1]}
\newcommand{\latency}[1]{\mathcal{L}(#1)}

\title{Increasing the Speed of Polar List Decoders}

\author{%
  Gabi Sarkis, %
  Pascal Giard,~\IEEEmembership{Student~Member,~IEEE}, %
  Alexander~Vardy,~\IEEEmembership{Fellow,~IEEE},\\%
  Claude~Thibeault,~\IEEEmembership{Senior~Member,~IEEE}, %
  and Warren~J.~Gross,~\IEEEmembership{Senior~Member,~IEEE}%
  \thanks{G. Sarkis, P. Giard, and W. J. Gross are with the Department of Electrical and Computer Engineering, McGill University, Montr\'eal, Qu\'ebec, Canada (e-mail: \{gabi.sarkis, pascal.giard\}@mail.mcgill.ca, warren.gross@mcgill.ca).}%
  \thanks{A. Vardy is with the Department of Electrical Engineering, University of California San Diego, La Jolla, CA. USA (e-mail: avardy@ucsd.edu).}%
  \thanks{C. Thibeault is with the Department of Electrical Engineering, \'Ecole de technologie sup\'erieure, Montr\'eal, Qu\'ebec, Canada (e-mail: claude.thibeault@etsmtl.ca).}}

\ifpdf
\hypersetup{%
  pdfauthor={Gabi Sarkis, Pascal Giard, %
    Alexander Vardy, Claude Thibeault, and Warren J. Gross}%
  ,pdftitle={Increasing the Speed of Polar List Decoders}%
}
\fi

\begin{document}

\maketitle

\begin{abstract}
In this work, we present a simplified successive cancellation list decoder that uses a Chase-like decoding process to achieve a six time improvement in speed compared to successive cancellation list decoding while maintaining the same error-correction performance advantage over standard successive-cancellation polar decoders. We discuss the algorithm and detail the data structures and methods used to obtain this speed-up. We also propose an adaptive decoding algorithm that significantly improves the throughput while retaining the error-correction performance. Simulation results over the additive white Gaussian noise channel are provided and show that the proposed system is up to 16 times faster than an LDPC decoder of the same frame size, code rate, and similar error-correction performance, making it more suitable for use as a software decoding solution.
\end{abstract}

\section{Introduction}
\label{sec:introduction}
Polar codes provably achieve the symmetric channel capacity as the code length $N$ increases, when they are decoded with the low-complexity successive-cancellation (SC) decoding algorithm \cite{Arikan2009}. However, the error-correction performance of SC decoding of polar codes at moderate lengths is mediocre. List \cite{ListIdo} and stack decoding \cite{StackNiu} have been proposed to improve the error-correction performance without increasing code length.

To further improve the error-correction capability of polar codes, various concatenation schemes have been proposed \cite{ListCRCIdo, PCCRCNiu, PCRSSamsung}. The most successful one is a serial concatenation of a polar code (PC) with a cyclic redundancy check (CRC) code, where the latter is used as an outer code \cite{ListCRCIdo}. For a given length $N$, the resulting code is shown to match or exceed the error-correction performance of turbo \cite{PCCRCNiu} as well as low-density parity-check (LDPC) codes\cite{ListCRCIdo}.

The throughput of SC decoders is low due to the serial nature of the algorithm. This issue was resolved by the simplified successive cancellation (SSC) \cite{Alamdar-Yazdi2011} and the Fast-SSC \cite{Sarkis2014} decoding algorithms. The latter of which has fast hardware \cite{Sarkis2014} and software decoders \cite{Giard2014}. Since list decoders are dependent on SC decoders as their major components, their throughput is also very low and they would benefit from improvements to the SC decoders. However, the SSC-based algorithms are not directly applicable to list, and list-CRC, decoding because they present a single estimate of codewords; whereas list decoders require multiple candidates with soft-valued reliabilities.

In this work, we modify the SSC algorithm to present multiple candidate codewords using a Chase-decoding-like process and we present SSC-based list decoders that offer higher throughput (average decoding speed) and lower latency (worst case decoding time) than their SC-based counterpart.

It was shown in \cite{Giard2014} that, for software implementations, polar decoders were faster than LDPC decoders with equivalent error-correction performance despite the longer lengths required for polar codes. In this work we show that software list polar decoders are faster than equivalent-performance LDPC decoders at the same code lengths.

We start with a review of polar codes, list and list-CRC decoding, and SSC decoding in Section~\ref{sec:background}. We present our SSC-List decoder in Section~\ref{sec:algorithm} and a higher throughput adaptive version in Section~\ref{sec:algo:adaptive}. Finally, we discuss the proposed decoder's throughput, latency, and error-correction performance in Section~\ref{sec:results}, comparing it with SC-List and LDPC decoders.

\section{Background}
\label{sec:background}

\subsection{Polar Codes}
\label{sec:bg:codes}
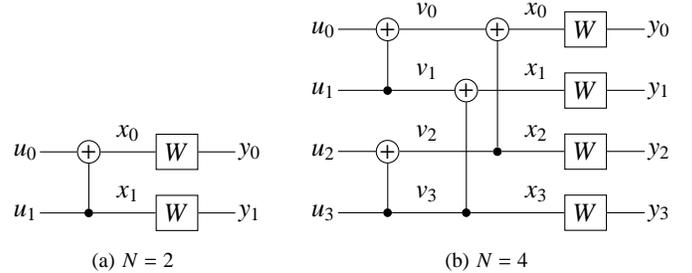
\begin{figure}[t]
  \centering
  \subfloat[$N=2$]{\label{fig:pc2}\newcommand{\ubit}[1]{$u_{#1}$}
\newcommand{\fbit}[1]{\color{gray}$u_{#1}$}
\begin{tikzpicture}

\usetikzlibrary{shapes,positioning,arrows,decorations.markings,fit}

\definecolor{varnode_fill}{RGB}{0,0,0}
\definecolor{chknode_fill}{RGB}{255,255,255}

\tikzset{
  chknode/.style={draw,fill=chknode_fill,circle,minimum size=0.3cm, inner sep=0},
  varnode/.style={draw,fill=varnode_fill,circle,minimum size=0.1cm, inner sep=0},
  channel/.style={draw,fill=white,rectangle},
  sep/.style={rectangle,minimum width=0.3cm, inner sep=0},
  bit/.style={circle, inner sep = 0}
}

\matrix[row sep=1mm, column sep=1mm] {
 	\node[bit] (n0s0) {\ubit{0}}; & \node[sep] {}; & \node[chknode] (n0s1) {$+$}; &  \node[sep,label={$x_0$}] {}; & \node[channel] (n0s2) {$W$}; & \node[sep] {}; & \node[bit] (n0s3) {$y_0$};\\
 	\node[bit] (n1s0) {\ubit{1}}; & \node[sep] {}; & \node[varnode] (n1s1) {}; &  \node[sep,label={$x_1$}] {}; & \node[channel] (n1s2) {$W$}; & \node[sep] {}; & \node[bit] (n1s3) {$y_1$};\\
};

\path[-] (n0s0) edge (n0s1);
\path[-] (n0s1) edge (n0s2);
\path[-] (n0s2) edge (n0s3);

\path[-] (n1s0) edge (n1s1);
\path[-] (n1s1) edge (n1s2);
\path[-] (n1s2) edge (n1s3);

\path[-] (n1s1) edge (n0s1);

\end{tikzpicture}}
  \subfloat[$N=4$]{\label{fig:pc4}\newcommand{\ubit}[1]{$u_{#1}$}
\newcommand{\fbit}[1]{\color{gray}$u_{#1}$}
\begin{tikzpicture}

\usetikzlibrary{shapes,positioning,arrows,decorations.markings,fit}

\definecolor{varnode_fill}{RGB}{0,0,0}
\definecolor{chknode_fill}{RGB}{255,255,255}

\tikzset{
  chknode/.style={draw,fill=chknode_fill,circle,minimum size=0.3cm, inner sep=0},
  varnode/.style={draw,fill=varnode_fill,circle,minimum size=0.1cm, inner sep=0},
  channel/.style={draw,fill=white,rectangle},
  sep/.style={rectangle,minimum width=0.31cm, inner sep=0},
  bit/.style={circle, inner sep = 0}
}

\tikzset{blue dotted/.style={draw=blue!50!white, line width=1pt,
    dash pattern=on 4pt off 4pt,
    inner sep=0.5mm, rectangle, rounded corners}};

\tikzset{blue dotted tight/.style={draw=blue!50!white, line width=1pt,
    dash pattern=on 4pt off 4pt,
    inner sep=0mm, rectangle, rounded corners}};

\matrix[row sep=1mm, column sep=1mm] {
  \node[bit] (n0u) {\ubit{0}}; & \node[sep] {}; & \node[chknode] (n0s1) {$+$}; & \node[sep,label=$v_0$] {}; && \node[chknode] (n0s2) {$+$}; &  \node[sep,label={$x_0$}] {}; & \node[channel] (n0c) {$W$}; & \node[sep] {}; & \node[bit] (n0y) {$y_0$};\\
  \node[bit] (n1u) {\ubit{1}}; & \node[sep] {}; & \node[varnode] (n1s1) {};    & \node[sep,label=$v_1$] {}; &  \node[chknode] (n1s2) {$+$}; && \node[sep,label={$x_1$}] {}; & \node[channel] (n1c) {$W$}; & \node[sep] {}; & \node[bit] (n1y) {$y_1$};\\
  \node[bit] (n2u) {\ubit{2}}; & \node[sep] {}; & \node[chknode] (n2s1) {$+$}; & \node[sep,label=$v_2$] {}; && \node[varnode] (n2s2) {};    &  \node[sep,label={$x_2$}] {}; & \node[channel] (n2c) {$W$}; & \node[sep] {}; & \node[bit] (n2y) {$y_2$};\\
  \node[bit] (n3u) {\ubit{3}}; & \node[sep] {}; & \node[varnode] (n3s1) {};    & \node[sep,label=$v_3$] {}; &  \node[varnode] (n3s2) {};    && \node[sep,label={$x_3$}] {}; & \node[channel] (n3c) {$W$}; & \node[sep] {}; & \node[bit] (n3y) {$y_3$};\\
};

\path[-] (n0u) edge (n0s1);
\path[-] (n0s1) edge (n0s2);
\path[-] (n0s2) edge (n0c);
\path[-] (n0c) edge (n0y);

\path[-] (n1u) edge (n1s1);
\path[-] (n1s1) edge (n1s2);
\path[-] (n1s2) edge (n1c);
\path[-] (n1c) edge (n1y);

\path[-] (n1s1) edge (n0s1);

\path[-] (n2u) edge (n2s1);
\path[-] (n2s1) edge (n2s2);
\path[-] (n2s2) edge (n2c);
\path[-] (n2c) edge (n2y);

\path[-] (n3u) edge (n3s1);
\path[-] (n3s1) edge (n3s2);
\path[-] (n3s2) edge (n3c);
\path[-] (n3c) edge (n3y);

\path[-] (n3s1) edge (n2s1);

\path[-] (n3s2) edge (n1s2);
\path[-] (n2s2) edge (n0s2);

\end{tikzpicture}}
  \caption{Construction of polar codes of lengths 2 and 4}
  \label{fig:construction}
\end{figure}

Polar codes approach the symmetric capacity of a channel $W$, as the code length $N \rightarrow \infty$, by exploiting channel polarization. Such constructions for $N\in\left\{2,4 \right\}$ are shown in Fig.~\ref{fig:construction}. In Fig.~\ref{fig:pc2}, the probability of correctly estimating $u_0$ given $y_0$ and $y_1$ is lower than that of correctly estimating $x_0$ given $y_0$, which is in turn lower than that of estimating $u_1$ given $y_0$, $y_1$, and $u_0$. Longer codes are built by recursively applying the linear polarizing construction. Fig.~\ref{fig:pc4} shows the case of $N=4$. As the code length increases, the probability of estimating each bit tends to either 0.5 (completely unreliable) or 1 (perfectly reliable). The proportion of the latter bits, called \emph{reliable} bits, approaches the capacity of the channel $W$ as $N \rightarrow \infty$ \cite{Arikan2009}.

To build an $(N,k)$ polar code, the $k$ information bits are transmitted through the $k$ most reliable locations. The remaining $N-k$ locations correspond to the least reliable bits and are set to 0 and called the \emph{frozen} bits. Determining the reliability of the bit locations depends on the type and conditions of the channel $W$ and is studied for different channels in \cite{Arikan2009} and \cite{Tal2011a}. A polar code is constructed for a given channel and channel condition, and can be represented using a generator matrix, $G_N = F_N = F_2^{\otimes \log_2 N}$, where $F_2 = \left[ \begin{smallmatrix} 1 & 0 \\ 1 & 1\end{smallmatrix} \right]$ and $^{\otimes}$ is the Kronecker power.  In \cite{Arikan2009}, a bit-reversal operator was used so that $G_N = B_N F_N$; however, it was shown in \cite{Giard2014} that not bit-reversing the rows of $F_N$ provides better memory layout and vectorization opportunities for software polar decoders. The frozen bits are indicated by setting their values to 0 in the source vector $\mvec{u}$.

SC decoding works sequentially by estimating an information (non-frozen) bit $u_i$ using the received channel values $\mvec{y}$ and the previously estimated bits $\estvec{0}{i - 1}$ according to
\begin{equation}
\label{eq:sc}
\est{i} = \begin{cases}
  0 & \text{when } \prob{\mvec{y}, \estvec{0}{i-1} | \est{i} = 0} \geq \prob{\mvec{y}, \estvec{0}{i-1} | \est{i} = 1};\\
  1 & \text{otherwise}.
\end{cases}
\end{equation}

\subsection{List-CRC Decoding}
\label{sec:bg:list-crc}
Instead of selecting one value for an estimate \eqref{eq:sc}, a list decoder works by assuming both 0 and 1 are estimates of the bit $u_i$ and generates two paths that are decoded using SC decoding. Without a set limit, the number of paths doubles for every information bit, growing exponentially and thus becoming a maximum-likelihood (ML) decoder. To constrain the complexity, a maximum of $L$ distinct paths, the most likely ones, are kept at the end of every step. Thus, a list decoder presents the $L$ most likely codeword candidates after it has estimated all bits. The codeword among the $L$ with the best path reliability metric, i.e. the largest likelihood value, is chosen to be the decoder output.

Noticing that when a polar list decoder failed, the correct codeword was often among the $L$ final candidates, the authors of \cite{ListIdo} proposed concatenating a CRC with the information bits, increasing the rate of the polar code to accommodate the additional bits and maintain the overall system rate. The CRC provides the criterion for selection from among the candidate, final codewords. The likelihood of the codewords is only consulted either when two or more candidates satisfy the CRC constraint or when none do. The resulting list-CRC decoder offers a significant improvement in error-correction performance over regular list decoding, to the extent where polar codes were shown to be able to outperform turbo codes\cite{PCCRCNiu} and LDPC codes\cite{ListIdo} of similar lengths and rates.

List-SC decoding, like SC decoding, remains a sequential process. Moreover, $L$ paths must now be decoded instead of one, increasing the latency from $O(N \log N)$ to $O(L N \log N)$ and decreasing throughput by the same factor \cite{ListIdo}.

To improve the decoder throughput, adaptive list decoding \cite{PCCRCLi} starts with $L=1$ and restarts with $L=2$ if the CRC is not satisfied. The list size is subsequently doubled until the constraint is satisfied or a maximum size, $L_{\text{max}}$, is reached, in which case the candidate with the highest reliability is selected. However, this method significantly increases latency, which becomes
\[
\latency{\text{A-SC-List}(L_{\text{max}})} = \sum_{l = 0}^{\log_2 L_{\text{max}} - 1} \latency{\text{SC-List}(2^l)};
\]
where $\text{A-SC-List}(L_{\text{max}})$ is an adaptive list decoder with a maximum list size of $L_{\text{max}}$ and $\text{SC-List}(L)$ is a list decoder with list size $L$.

\subsection{SSC Decoding}
\label{sec:bg:ssc}
\begin{figure}[t]
  \centering
  \subfloat[SC]{\label{fig:sc-tree}\begin{tikzpicture}[baseline = (0_7.center),
        level/.style={level distance = 6mm},
        level 1/.style={sibling distance=19mm, edge from parent/.style={draw,black,line width=2pt}},
        level 2/.style={level distance=10mm, sibling distance=9.5mm, edge from parent/.style={draw,black,line width=1pt}},
        level 3/.style={sibling distance=4.7mm, edge from parent/.style={draw,black,line width=0.5pt}},
        ]

\tikzset{
frozen/.style={thick,draw=black,fill=white,minimum size=3mm,circle, inner sep=0},
fullspace/.style={thick,draw=black,fill=black,minimum size=3mm,circle, inner sep = 0},
mixed/.style={thick,draw=black,fill=gray,minimum size=3mm,circle, inner sep = 0},
ml_mixed/.style={thick,draw=black,fill=blue,minimum size=3mm,circle, inner sep = 0}
}

\node[mixed] (p){} [grow=left]
	child {node[frozen] (2_0){}
		child {node[frozen] (1_0){}
			child {node[frozen] (a0_0){}
			}
			child {node[frozen] (a0_1){}
			}
		}
		child {node[frozen] (1_2){}
			child {node[frozen] (0_2){}
			}
			child {node[frozen] (0_3){}
			}
		}
	}
	child {node[mixed] (v){}
		child {node[mixed] (cl){}
			child {node[frozen] (0_4){}
			}
			child {node[fullspace] (0_5){}
			}
		}
		child {node[fullspace] (cr){}
			child {node[fullspace] (0_6){}
			}
			child {node[fullspace] (0_7){}
			}
		}
	}
;




\end{tikzpicture}}
  \quad
  \subfloat[SSC]{\label{fig:ssc-tree} \begin{tikzpicture}[baseline=(base.center),
        level/.style={level distance = 6mm},
        level 1/.style={sibling distance=19mm, edge from parent/.style={draw,black,line width=2pt}},
        level 2/.style={sibling distance=9mm, edge from parent/.style={draw,black,line width=1pt}},
        level 3/.style={sibling distance=4mm, edge from parent/.style={draw,black,line width=0.5pt}},
        ]

\tikzset{
frozen/.style={thick,draw=black,fill=white,minimum size=3mm,circle, inner sep=0},
fullspace/.style={thick,draw=black,fill=black,minimum size=3mm,circle, inner sep = 0},
mixed/.style={thick,draw=black,fill=gray,minimum size=3mm,circle, inner sep = 0},
ml_mixed/.style={thick,draw=black,fill=blue,minimum size=3mm,circle, inner sep = 0}
}

\node[mixed] (3_0){} [grow=left]
	child {node[frozen] (2_0){}
	}
	child {node[mixed] (2_1){}
		child {node[mixed] (1_2){}
			child {node[frozen] (0_4){}
			}
			child {node[fullspace] (0_5){}
			}
		}
		child {node[fullspace] (1_3){}
		}
	}
;

\node [circle,below= 0.27mm of 1_3.base] (base) {};

\end{tikzpicture}}
  \caption{Decoder trees corresponding to the SC and SSC decoding algorithms}
\end{figure}
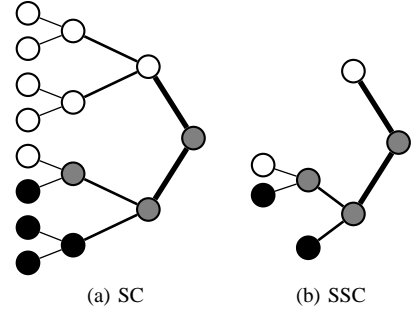

The recursive construction of a polar code makes binary trees a natural representation where each node corresponds to a constituent code of length $N_v$ with a soft input $\alpha$ and an estimated codeword output $\beta$. It was observed in \cite{Alamdar-Yazdi2011} that a subtree where all leaf-nodes correspond to frozen bits need not be traversed; its output is known a priori to be the zero-vector. Similarly, that work showed that the ML output of a subtree where all leaf-nodes are information bits, i.e. corresponding to constituent code of rate 1, can be obtained by performing threshold detection on the soft-information input vector. Any tree corresponding to a rate $R$ code is traversed until a rate-0 or a rate-1 code is reached. As a result of these observations, the decoder tree is pruned resulting in the simplified SC (SSC) decoder tree. Decoder trees for the SC and SSC algorithms decoding the same code are shown in Fig.~\ref{fig:sc-tree} and Fig.~\ref{fig:ssc-tree} respectively. For the SC decoder tree, white leaves correspond to frozen bits and black leaves correspond to information bits. For the SSC decoder tree, white and black leaves are called rate-0 and rate-1 nodes respectively.

The decoder tree is further pruned in \cite{Sarkis2013,Sarkis2014} by recognizing more types of constituent codes, resulting in lower latency and greater throughput for both hardware\cite{Sarkis2014} and software decoders\cite{Giard2014}.

\section{SSC-List Decoder}
\label{sec:algorithm}
In this section, we present an SSC-based list decoding algorithm and discuss its implementation details.
Rate-0 nodes are ignored and their soft-input is not calculated by their parent, and rate-R nodes operate as in SC-List decoding. Therefore we focus on rate-1 nodes. We will show in Section~\ref{sec:vs-sc} that the proposed decoder is six times as fast the SC-List decoder.

It should be noted that this decoder was implemented using log-likelihoods (LL) to represent bit reliabilities.

\subsection{Chase-Like Decoding of Rate-1 Nodes}
The function of the rate-1 node decoder is to provide a list of the $L$ most reliable candidate codewords given its LL input $\mvec{\alpha}$, where each LL $\alpha[i]$ consists of $\alpha_0[i]$ and $\alpha_1[i]$. For a constituent code of rate 1 and length $N_v$, there exists $2^{N_v}$ candidate codewords, rendering an exhaustive search impractical for all but the smallest of such codes. Therefore, we employ the candidate generation method of Chase decoding \cite{Chase1972}.

Maximum-likelihood decoding of a rate-1 constituent code is performed on a bit-by-bit basis \cite{Alamdar-Yazdi2011}, i.e.
\[
\beta[i] = \begin{cases}
  0 & \text{when } \alpha_0[i] \geq \alpha_1[i],\\
  1 & \text{otherwise.}
\end{cases}
\]
To provide a list of candidate codewords, the least reliable bits---determined using $r[i] = |\alpha_0[i] - \alpha_1[i]|$---of the ML decision are flipped individually.
Simulation results have shown that two-bit errors must also be considered. Therefore, the list of candidates is augmented with codewords that differ from the ML decision by two of the least reliable bits.

The list of candidates is pruned to include, at most, $L$ candidates. This is accomplished by discarding the least reliable candidates, where the reliability of a path $x$ with an estimated output $\beta$ is calculated according to
\begin{equation}
R_x = \sum_i \alpha_{\beta[i]}[i].
\end{equation}

\subsection{Implementation of Rate-1 Decoders}
The rate-1 decoder starts by initializing its set of candidates to an empty set. Then, for each source path $p$, it will calculate and store the ML decision and generate a set of candidate forks. Once the decoder has iterated over all source paths, it will store the up to $L$ most reliable paths from the ML decisions and the candidate forks, discarding the rest. The top-level function corresponds to Algorithm~\ref{algo:r1}. The algorithm shows how the bit reliabilities $r$ and the path reliability $R$ are calculated in tandem with the ML decision. The candidate forks are appended to the candidate set when there are fewer than $L$ candidates already stored; otherwise, they replace other candidates with lower reliability.

\begin{algorithm}[t]
  \caption{decodeRate1Code}
  \label{algo:r1}
  \begin{algorithmic}[1]
    \State $\textsl{candidates} = \{\}$
    \For {$p \in \textsl{sourcePaths}$}
    \State $R_p = 0$
    \For {$i =0 \textbf{ to } N_v-1$}
    \State $\beta^p[i] = \argmax_x( \alpha_x^p[i])$
    \State $r[i] = |\alpha_0^p[i] - \alpha_1^p[i]|$
    \State $R_p = R_p + \max(\alpha_0^p[i], \alpha_1^p[i])$
    \EndFor
    \State storePath($p$, $R_p$)
    \If {$\textsl{candidates.count} < L$}
    \State appendCandidates(\textsl{candidates})
    \Else
    \State replaceCandidates(\textsl{candidates})
    \EndIf
    \EndFor
    \State mergeBestCandidates(\textsl{candidates})
  \end{algorithmic}
\end{algorithm}

Algorithm~\ref{algo:append}, shows how candidates are appended to the set. Empirically, it was observed that not all bits need to be considered when enumerating potential single-bit errors, limiting the search to the $c$ least reliable bits was sufficient, as in Chase decoding \cite{Chase1972}. Therefore, this method performs a partial sort to find those bits. The candidates are generated by flipping those bits individually, and their reliabilities are calculated according to 
\begin{align}
R_i &= R_p - r[i] = R_p - |\alpha_0^p[i] - \alpha_1^p[i]|\nonumber\\
&= R_p - \max(\alpha_0^p[i], \alpha_1^p[i]) + \min(\alpha_0^p[i], \alpha_1^p[i]).\nonumber
\end{align}
Since a candidate might be later discarded if it is not among the $L$ most reliable paths, it is important for speed reasons to minimize the amount of information stored about each candidate. Therefore only the information needed to construct a new path is stored in the candidate set: the source path $p$, the path reliability $R_i$, and the location of bits in which it differs from the source path \textsl{bitsToFlip}. Candidates with two-bit errors are generated in a similar manner by iterating over all unique pairs of bits among the $c$ least reliable ones.
To remove conditionals from the inner loops in this algorithm, the set of candidates is allowed to contain more than $L$ candidates. Selecting the correct number of candidates to store as new paths, is performed at a later point by the rate-1 decoder.

\begin{algorithm}[t]
  \caption{appendCandidates}
  \label{algo:append}
  \begin{algorithmic}[1]
    \State\Comment{Appends forks of path $p$ to \textsl{candidates} with constraint $c$}

    \State partialSort($r$, $c$)

    \For {$i = 0 \textbf{ to } c-1$} \Comment{Single-bit errors}
    \State $R_i = R_p - r[i]$
    \State $\textsl{bitsToFlip} = \{\text{bitIndex}(i)\}$
    \State \textsl{candidates}.insert($p$, $R_i$, \textsl{bitsToFlip})
    \EndFor

    \For {$i = 0 \textbf{ to } c-2$} \Comment{Two-bit errors}
    \For {$j = i+1 \textbf{ to } c-1$}
    \State $R_{ij} = R_p - r[i] - r[j]$
    \State $\textsl{bitsToFlip} = \{\text{bitIndex}(i),\; \text{bitIndex}(j)\}$
    \State \textsl{candidates}.insert($p$, $R_{ij}$, \textsl{bitsToFlip})
    \EndFor
    \EndFor
  \end{algorithmic}
\end{algorithm}

When the set of candidates already contains $L$ or more candidates, the decoder will only replace an existing candidate with a new one when the latter is more reliable. Algorithm~\ref{algo:replace} describes this process. It iterates over candidates with single-bit and two-bit errors and adds them to the set of candidates if their reliability is greater than the minimum stored in the set. Every time a new candidate is added to the set, the least reliable one is removed. This prevents the set of candidates from storing a large number of candidates that will be discarded later. Similar to Algorithm~\ref{algo:append}, it was observed via simulations that using a constraint $c$ to limit the candidate search space did not noticeably affect error-correction performance while doubling the decoding speed.

Once the candidates for all sources paths have been generated, the most reliable $L$ of them are considered for use as paths replacing less reliable ML decisions of other paths if necessary. This is performed by the mergeBestCandidates() method where the new paths have their $\beta$ value stored by copying and modifying the ML decision of their source path.

In Algorithms \ref{algo:append} and \ref{algo:replace}, it is observed that the most common operations performed on the set of candidates, denoted \textsl{candidates}, are insertion, deletion, and finding the minimum. Red-Black trees are well suited for implementing such a data structure since all these operations are performed in $O(\log_2 N_v)$ time in the worst case \cite{Cormen2009}.
In addition, mergeBestCandidates() requires that the most reliable candidates be indicated and red-black trees store their contents sorted by key.

\begin{algorithm}[t]
  \caption{replaceCandidates}
  \label{algo:replace}
  \begin{algorithmic}[1]
    \State\Comment{Replaces the least reliable \textsl{candidates} with more reliable forks of path $p$}.

    \State partialSort($r$, $c$)

    \For {$i = 0 \textbf{ to } c-1$} \Comment{Single-bit errors}
    \State $R_i = R_p - r[i]$
    \If {$R_i > \min (\textsl{candidates}.reliability)$}
    \State $\textsl{bitsToFlip} = \{\text{bitIndex}(i)\}$
    \State \textsl{candidates}.insert($p$, $R_i$, \textsl{bitsToFlip})
    \State \textsl{candidates}.remove(\textsl{candidates}.leastReliable)
    \EndIf
    \EndFor

    \For {$i = 0 \textbf{ to } c-2$} \Comment{Two-bit errors}
    \For {$j = i+1 \textbf{ to } c-1$}
    \State $R_{ij} = R_p - r[i] - r[j]$
    \If {$R_{ij} > \min (\textsl{candidates}.reliability)$}
    \State $\textsl{bitsToFlip} = \{\text{bitIndex}(i),\; \text{bitIndex}(j)\}$
    \State \textsl{candidates}.insert($p$, $R_{ij}$, \textsl{bitsToFlip})
    \State \textsl{candidates}.remove(\textsl{candidates}.leastReliable)
    \EndIf
    \EndFor
    \EndFor
  \end{algorithmic}
\end{algorithm}

\section{Adaptive SSC-List-CRC Decoder}
\label{sec:algo:adaptive}
List decoders have a high latency and a low throughput that are constant regardless of the channel condition. Based on the observation that at high $\nicefrac{E_b}{N_0}$ values the average list size $L$ required to successfully correct a frame is low, an adaptive SC-List-CRC decoder was proposed in \cite{PCCRCLi}.

In Section~\ref{sec:algorithm}, we introduced an SSC-List decoding algorithm that has a lower latency and greater throughput than the SC-List decoding algorithm. Despite the improvement, the throughput of that decoder is still significantly lower than a Fast-SSC decoder \cite{Sarkis2014}. We thus propose using an adaptive SSC-List-CRC decoding algorithm similar to that of \cite{PCCRCLi}:
\begin{enumerate}
\item Decode a frame using the Fast-SSC algorithm.
\item Verify the validity of the estimated codeword by calculating its CRC.
\item Stop the decoding process if the CRC is satisfied, otherwise move to the next step.
\item Relaunch the decoding process using the SSC-List algorithm and generate a list of $L$ candidate codewords sorted by their path reliability metric.
\item Pick the most reliable candidate among the list generated above that satisfies the CRC.
\item If none of the $L$ candidates satisfy the CRC, pick the codeword with the best path reliability metric.
\end{enumerate}

The difference between this proposed algorithm and that of \cite{PCCRCLi} is that in order to reduce latency, the list size is not increased gradually. Instead, it is changed from $L=1$, i.e. using the Fast-SSC decoder, to $L=L_{\text{max}}$. Therefore, the latency (worst case) is
\begin{align}
&\latency{\text{A-SSC-List}(L_{\text{max}})}\nonumber\\
& = \latency{\text{SSC-List}(L_{\text{max}})} + \latency{\text{Fast-SSC}}\nonumber\\
& \approx \latency{\text{SSC-List}(L_{\text{max}})}.\nonumber
\end{align}
Since the latency of the single SSC-List decoder using $L = L_{\text{max}}$ is much greater than that of the Fast-SSC decoder.

Let $\latency{\text{L}} = \latency{\text{SSC-List}(L_{\text{max}})}$ and $\latency{\text{F}} = \latency{\text{Fast-SSC}}$, and denote the frame-error rate (FER) at the output of the Fast-SSC decoder $\text{FER}_{\text{F}}$. The expression for the information throughput (on average) of the proposed adaptive SSC-List decoder when decoding a code with dimension $k$ is
\[
\mathcal{T} = \frac{k}{(1 - \text{FER}_{\text{F}})\latency{F} + \text{FER}_{\text{F}}\latency{L}};
\]
where it can be observed that for sufficiently low $\text{FER}_{\text{F}}$ value, the throughput will determined mostly by the speed of the Fast-SSC decoder.

\section{Simulation Results}
\label{sec:results}

\subsection{Methodology}
All error-correction performance results were obtained for the binary-input additive white Gaussian noise (AWGN) channel with random codewords and binary phase-shift keying (BPSK) modulation. Polar codes were constructed using the technique described in \cite{Tal2011a} and systematic encoding was used \cite{Arikan2011}.
The throughput and latency values were measured on an Intel Core-i7 2600 running at 3.4 GHz using the methodology described in \cite{Giard2014}. Finally, as mentioned in Section~\ref{sec:bg:list-crc}, in list-CRC decoders, the rate of the polar code is adjusted to maintain the same overall system rate. For example, when comparing a list-CRC decoder with the (2048, 1723) LDPC decoder and a 32-bit CRC is utilized, the polar code used is PC(2048, 1755) and the overall system rate remains $1723/2048$.

\subsection{Choosing a Suitable CRC Length}
\begin{figure}[t]
 \centering
 \begin{tikzpicture}

  \pgfplotsset{
    grid style = {
      dash pattern = on 0.05mm off 1mm,
      line cap = round,
      black,
      line width = 0.5pt
    }
  }

  \begin{semilogyaxis}[%
      xlabel=$E_b/N_0$ (dB),%
      xlabel style={yshift=0.6em},%
      ylabel=FER, ylabel style={yshift=-0.65em},%
      width=0.5\columnwidth, height=8cm, grid=major,%
      legend style={
        anchor={center},
        cells={anchor=west},
        column sep=2mm,
        font=\footnotesize,
      },
      legend columns=3,%
      legend to name=crc-floor-legend,%
      mark size=3.0pt]

    \addplot[color=black,mark=star] table[x=snr_db,y=FER] {data/pc1024_860_l128_c08_listcrc.csv};
    \addlegendentry{$CRC=8$}

    \addplot[color=red,mark=triangle] table[x=snr_db,y=FER] {data/pc1024_860_l128_c32_listcrc.csv};
    \addlegendentry{$CRC=32$}

  \end{semilogyaxis}
\end{tikzpicture}
\begin{tikzpicture}

  \pgfplotsset{
    grid style = {
      dash pattern = on 0.05mm off 1mm,
      line cap = round,
      black,
      line width = 0.5pt
    }
  }

  \begin{semilogyaxis}[%
      xlabel=$E_b/N_0$ (dB),%
      xlabel style={yshift=0.6em},%
      ylabel=BER, ylabel style={yshift=-0.65em},%
      width=0.5\columnwidth, height=8cm, grid=major,%
      mark size=3.0pt]

    \addplot[color=black,mark=star] table[x=snr_db,y=BER] {data/pc1024_860_l128_c08_listcrc.csv};

    \addplot[color=red,mark=triangle] table[x=snr_db,y=BER] {data/pc1024_860_l128_c32_listcrc.csv};

  \end{semilogyaxis}
\end{tikzpicture}
\\
\ref{crc-floor-legend}
 \caption{The effect of CRC length on the error-correction performance of $(1024, 860)$ list-CRC decoders with $L=128$.}
 \label{fig:crc_floor}
\end{figure}
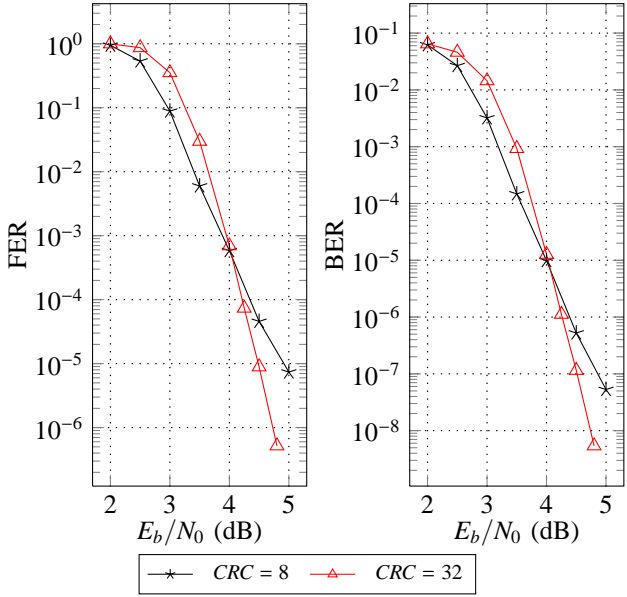
As discussed in Section~\ref{sec:bg:list-crc}, a CRC serves as better criterion for selecting the correct codeword from the final $L$ candidates even when its likelihood is not the largest. The length of the chosen CRC has an impact on the error-rate that varies with $\nicefrac{E_b}{N_0}$. Fig.~\ref{fig:crc_floor} shows the error-correction performance of a $(1024, 860)$ system consisting of polar code concatenated with a CRC of length 8 or 32 and decoded with a list-CRC decoder with list size $L=128$. It shows that a polar code concatenated with the shorter CRC will perform better at lower $\nicefrac{E_b}{N_0}$ values but will eventually achieve higher error-rates than the polar code concatenated with the longer CRC.

Therefore, the length of the CRC can be chosen to improve error-correction performance in the targeted $\nicefrac{E_b}{N_0}$ or BER/FER range.
\subsection{Error-Correction Performance}

\begin{figure}[t]
 \centering
 \begin{tikzpicture}

  \pgfplotsset{
    grid style = {
      dash pattern = on 0.05mm off 1mm,
      line cap = round,
      black,
      line width = 0.5pt
    }
  }

  \begin{semilogyaxis}[%
    xlabel=$E_b/N_0$ (dB),%
    xlabel style={yshift=0.6em},%
    ylabel=FER, ylabel style={yshift=-0.65em},%
    width=0.5\columnwidth, height=8cm, grid=major,%
    legend style={
      anchor={center},
      cells={anchor=west},
      column sep= 2mm,
      font=\footnotesize,
    },
    legend to name=perf-list-legend,
    legend columns=3,
    mark size=3.0pt]

    \addplot[color=blue,mark=pentagon] table[x=snr_db,y=FER] {data/pc-a2k.o7724};
    \addlegendentry{SC-List-CRC}

    \addplot[color=red,mark=triangle] table[x=snr_db,y=FER] {data/list-ssc-a2k32.o109};
    \addlegendentry{SSC-List-CRC}

  \end{semilogyaxis}
\end{tikzpicture}
\begin{tikzpicture}

  \pgfplotsset{
    grid style = {
      dash pattern = on 0.05mm off 1mm,
      line cap = round,
      black,
      line width = 0.5pt
    }
  }

  \begin{semilogyaxis}[%
    xlabel=$E_b/N_0$ (dB),%
    xlabel style={yshift=0.6em},%
    ylabel=BER, ylabel style={yshift=-0.65em},%
    width=0.5\columnwidth, height=8cm, grid=major,%
    mark size=3.0pt]

    \addplot[color=blue,mark=pentagon] table[x=snr_db,y=BER] {data/pc-a2k.o7724};

    \addplot[color=red,mark=triangle] table[x=snr_db,y=BER] {data/list-ssc-a2k32.o109};

  \end{semilogyaxis}
\end{tikzpicture}
\\
\ref{perf-list-legend}
 \caption{Error-correction performance of $(2048,1723)$ SC- and SSC-List-CRC decoders with $L=32$.}
 \label{fig:l32}
\end{figure}
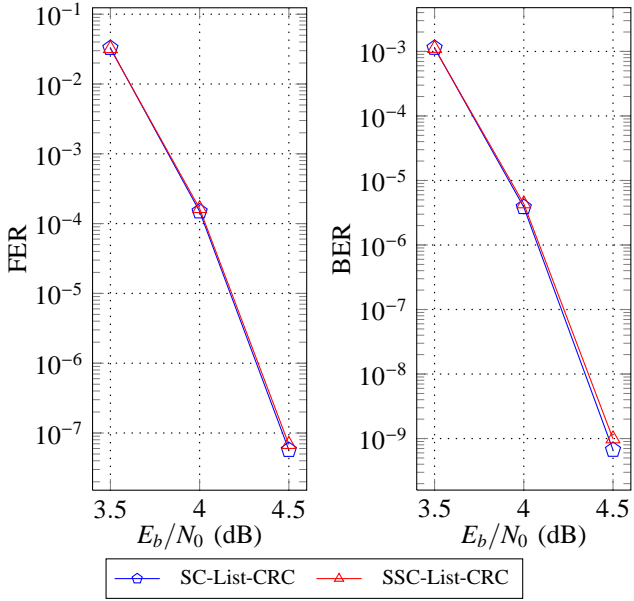

It is known that concatenating CRC improves the error-correction performance of polar list decoders. In this section, we first show that the error-correction performance of the proposed SSC-List-CRC decoder is the same as that of the SC-List-CRC decoder in Fig.~\ref{fig:l32}. We then demonstrate that the benefits for longer codes are still significant.

\begin{figure}[t]
 \centering
 \begin{tikzpicture}

  \pgfplotsset{
    grid style = {
      dash pattern = on 0.05mm off 1mm,
      line cap = round,
      black,
      line width = 0.5pt
    }
  }

  \begin{semilogyaxis}[%
    xlabel=$E_b/N_0$ (dB),%
    xlabel style={yshift=0.6em},%
    ylabel=FER, ylabel style={yshift=-0.65em},%
    width=0.5\columnwidth, height=8cm, grid=major,%
    legend style={
      anchor={center},
      cells={anchor=west},
      column sep= 2mm,
      font=\footnotesize,
    },
    legend to name=perf-legend,
    legend columns=3,
    mark size=3.0pt]

    \addplot[color=blue,mark=pentagon] table[x=snr_db,y=FER] {data/mlssc32k9.o8134};
    \addlegendentry{PC}

    \addplot[color=red,mark=triangle] table[x=snr_db,y=FER] {data/fl32k-l4x.o80};
    \addlegendentry{List-CRC $L=4$}

    \addplot[color=black,mark=asterisk] table[x=snr_db,y=FER] {data/fl32k-l32x.o86};
    \addlegendentry{List-CRC $L=32$}

  \end{semilogyaxis}
\end{tikzpicture}
\begin{tikzpicture}

  \pgfplotsset{
    grid style = {
      dash pattern = on 0.05mm off 1mm,
      line cap = round,
      black,
      line width = 0.5pt
    }
  }

  \begin{semilogyaxis}[%
    xlabel=$E_b/N_0$ (dB),%
    xlabel style={yshift=0.6em},%
    ylabel=BER, ylabel style={yshift=-0.65em},%
    width=0.5\columnwidth, height=8cm, grid=major,%
    mark size=3.0pt]

    \addplot[color=black,mark=asterisk] table[x=snr_db,y=BER] {data/fl32k-l32x.o86};

    \addplot[color=red,mark=triangle] table[x=snr_db,y=BER] {data/fl32k-l4x.o80};

    \addplot[color=blue,mark=pentagon] table[x=snr_db,y=BER] {data/mlssc32k9.o8134};

  \end{semilogyaxis}
\end{tikzpicture}
\\
\ref{perf-legend}
 \caption{Error-correction performance of (32768, 29492) polar code, denoted PC, with that of (32768, 29492) list-CRC decoders with different list sizes.}
 \label{fig:ecc_perf_32k}
\end{figure}
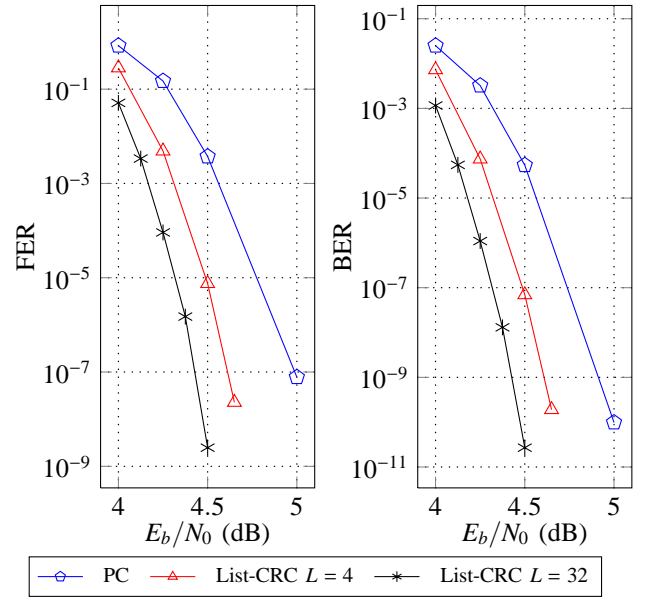

As shown in Fig.~\ref{fig:ecc_perf_32k}, for a $(32768, 29492)$ polar code, the use of the proposed algorithm results in a coding gain greater than 0.3 dB and 0.5 dB at a FER of $10^{-5}$ over the Fast-SSC algorithm for $L=4$ and $L=32$, respectively. It can be seen that the curves are diverging as $\nicefrac{E_b}{N_0}$ is increasing, and thus the coding gain is growing as well.

\subsection{Comparison with the SC-List-CRC Decoder}
\label{sec:vs-sc}
List decoders have latency and throughput that are constant across $\nicefrac{E_b}{N_0}$ values. Table~\ref{tab:list-tp} shows these values for the SC-List-CRC and SSC-List-CRC decoders for two different list sizes when decoding a (2048, 1723) polar+CRC-32 code. At $L=32$, the SSC-based decoder is approximately 7 times as fast the SC-based one. At $L=128$, it is 6 times as fast.

\begin{table}[t]
  \caption{Latency and information throughput comparison for list-based decoders using a $(2048,1723)$ polar+CRC code.}
  \centering
  \begin{tabular}{r c c c}
    \toprule
    Decoder & $L$ & Latency (ms) & $\mathcal{T}$ (kbps)\\
    \midrule
    SC-List-CRC     & 32  & 23 & 74\\
    SSC-List-CRC    & & 3.3 &  522\vspace{2pt}\\
    SC-List-CRC     & 128 & 97 & 17\\
    SSC-List-CRC    & & 16 & 107\\
    \bottomrule
  \end{tabular}
  \label{tab:list-tp}
\end{table}

\subsection{Comparison with LDPC Codes}
To the best of our knowledge, the fastest CPU-based LDPC decoder published in literature is that of \cite{Falcao2011}. Its information throughput for a (1024, 512) LDPC running on two CPU cores was 345 kbps with a fixed number of iterations (10). The information throughput of a scaled-min-sum decoder we have developed was 555 kbps when running with the same number of iterations but on a single CPU core of similar speed. Therefore, we use our LDPC decoder for throughput comparison in this work and enable early termination to further improve its throughput.

\begin{figure}[t]
 \centering
 \begin{tikzpicture}

  \pgfplotsset{
    grid style = {
      dash pattern = on 0.05mm off 1mm,
      line cap = round,
      black,
      line width = 0.5pt
    }
  }

  \begin{semilogyaxis}[%
    xlabel=$E_b/N_0$ (dB),%
    xlabel style={yshift=0.6em},%
    ylabel=FER, ylabel style={yshift=-0.65em},%
    xmin=3.25,
    xmax=5,
    width=0.5\columnwidth, height=8cm, grid=major,%
    legend style={
      anchor={center},
      cells={anchor=west},
      column sep= 2mm,
      font=\footnotesize,
    },
    legend columns=2,%
    legend to name=ldpc-perf-legend,
    mark size=3.0pt]

    \addplot[color=black,mark=asterisk] table[x=snr_db,y=FER] {data/10g.o114};
    \addlegendentry{LDPC, $I_{\max} = 30$}

    \addplot[color=red,mark=triangle] table[x=snr_db,y=FER] {data/list-ssc-a2k32.o109};
    \addlegendentry{PC List-CRC, $L=32$}

    \addplot[color=blue,mark=square] table[x=snr_db,y=FER] {data/list-ssc-a2k32.o113};
    \addlegendentry{PC List-CRC, $L=64$}

  \end{semilogyaxis}
\end{tikzpicture}
\begin{tikzpicture}

  \pgfplotsset{
    grid style = {
      dash pattern = on 0.05mm off 1mm,
      line cap = round,
      black,
      line width = 0.5pt
    }
  }

  \begin{semilogyaxis}[%
    xlabel=$E_b/N_0$ (dB),%
    xlabel style={yshift=0.6em},%
    ylabel=BER, ylabel style={yshift=-0.65em},%
    width=0.5\columnwidth, height=8cm, grid=major,%
    mark size=3.0pt]

    \addplot[color=black,mark=asterisk] table[x=snr_db,y=BER] {data/10g.o114};

    \addplot[color=red,mark=triangle] table[x=snr_db,y=BER] {data/list-ssc-a2k32.o109};

    \addplot[color=blue,mark=square] table[x=snr_db,y=BER] {data/list-ssc-a2k32.o113};

  \end{semilogyaxis}
\end{tikzpicture}
\\
\ref{ldpc-perf-legend}
 \caption{Error-correction performance of $(2048,1723)$ LDPC and polar list-CRC decoders.}
 \label{fig:ldpc}
\end{figure}
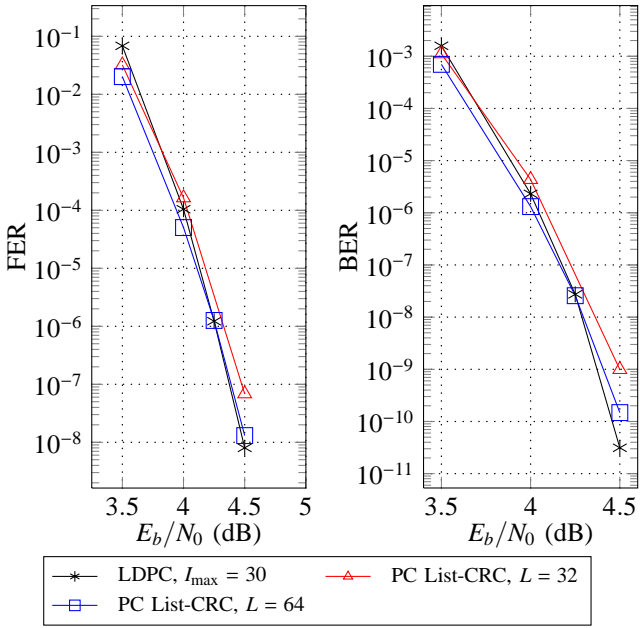

A polar list-CRC decoder with a 32-bit CRC and $L=32$ is within 0.1 dB of the error-correction performance of the 10GBASE-T (802.3an) LDPC code with identical code length and dimension (2048, 1723) as shown in Fig.~\ref{fig:ldpc}. When the list size is increased to 64, the polar list-CRC and the LDPC decoders have similar performance. In these simulations the LDPC decoder was using the scaled-min-sum algorithm with a maximum of 30 iterations ($I_{\max} = 30$) and a scaling factor of 0.5.

\begin{table}[t]
  \caption{Information throughput in Mbps of the proposed decoder compared to an LDPC decoder at different $\nicefrac{E_b}{N_0}$ values.}
  \centering
  \begin{tabular}{r c c c}
    \toprule
    \multirow{2}{*}{Decoder}& \multicolumn{3}{c}{$\mathcal{T}$ (Mbps)}\\
    \cmidrule{2-4}
    & 3.5 dB & 4.0 dB & 4.5 dB\\
    \midrule
    LDPC & 1.04 & 1.81 & 2.25\\
    A. SSC-List-CRC ($L = 64$) & 0.42 & 2.36 & 36.6\\
    A. SSC-List-CRC ($L = 32$) & 0.91 & 4.90 & 54.0\\
    \bottomrule
  \end{tabular}
  \label{tab:ldpc-tp}
\end{table}

Table~\ref{tab:ldpc-tp} shows the throughput values for the proposed adaptive SSC-List-CRC decoder with $L=64$ compared with that of our offset-min-sum LDPC decoder with $I_{\max} = 30$ and an adaptive SC-List-CRC decoder at different $\nicefrac{E_b}{N_0}$ values when decoding (2048, 1723) codes. We first observe that throughput of the decoders improves as $\nicefrac{E_b}{N_0}$ increases since they employ early termination methods: syndrome checking for the LDPC decoder and CRC checking for the adaptive SSC-List one. The LDPC decoder is faster than the proposed decoder at $\nicefrac{E_b}{N_0} = 3.5$ dB. At 4.0 dB and 4.5 dB however, the adaptive SSC-List decoder becomes 1.3 and 16 times as fast as the LDPC one, respectively. The latency was 5.5 ms and 7.1 ms for the LDPC and adaptive SSC-List decoders, respectively.
The table also shows the throughput of the adaptive SSC-List decoder with $L=32$, which at 3.5 dB runs at 87\% the speed of the LDPC decoder and is 2.7 and 24 times as fast at 4.0 dB and 4.5 dB, respectively. The latency of this decoder is 3.3 ms and, as mentioned in this section, its error-correction performance is within 0.1 dB of the LDPC decoder.

\section{Conclusion}
\label{sec:conclusion}
In this work, we presented a new polar list decoding algorithm whose software implementation is at least 6 times as fast as the original list decoder. We also showed an adaptive decoder which significantly increased the throughput to the point where its throughput is up to 16 times that of an LDPC decoder of the same length, rate, and similar error-correction performance. We believe that such improvements in speed, combined with the error-correction performance, make the adaptive SSC-List decoder a viable option for use as a decoder in software defined radio and other applications. Future work will focus on switching the list decoder to log-likelihood ratios in order to further reduce latency.

\bibliographystyle{IEEEtran}
\bibliography{IEEEabrv,refs.bib}

\end{document}